\documentclass{egpubl}

\usepackage{amsmath}
\usepackage{amssymb} %
\usepackage{booktabs}
 \usepackage{amsfonts}
 \usepackage{xurl}
\AtBeginDocument{\hypersetup{pdfkeywords={foveated rendering, motion judder, Gabor noise, perception}, pdfsubject={}}}
\copyrightTextTitPag{}
\copyrightTextRunPag{}
\makeatletter
\let\ps@titlepage\ps@plain
\makeatother

\usepackage[T1]{fontenc}
\usepackage{dfadobe}

\usepackage{cite}  %
\BibtexOrBiblatex
\electronicVersion
\PrintedOrElectronic
\ifpdf \usepackage[pdftex]{graphicx} \pdfcompresslevel=9
\else \usepackage[dvips]{graphicx} \fi

\usepackage{egweblnk}
\usepackage{xcolor}
\usepackage{subcaption}

\usepackage{setspace}

\author[V. Cantory and D. Zhang and V. Interrante and P. Didyk]{%
  Ville Cantory$^{1,\dagger}$,
  Danhua Zhang$^{1,*}$,
  Victoria Interrante$^{1,\ddagger}$,
  and Piotr Didyk$^{2,\S}$\\
  $^{1}$University of Minnesota, USA\\
  $^{2}$Università della Svizzera italiana, Lugano, Switzerland\\
  {\footnotesize
    \mbox{$^{\dagger}$\href{mailto:canto063@umn.edu}{\nolinkurl{canto063@umn.edu}}}\quad
    \mbox{$^{*}$\href{mailto:zhan5954@umn.edu}{\nolinkurl{zhan5954@umn.edu}}}}\\
  {\footnotesize
    \mbox{$^{\ddagger}$\href{mailto:interran@umn.edu}{\nolinkurl{interran@umn.edu}}}\quad
    \mbox{$^{\S}$\href{mailto:piotr.didyk@usi.ch}{\nolinkurl{piotr.didyk@usi.ch}}}}
}

\title[Masking Judder in Foveated Rendering with Phase-Shifted Noise]%
      {Masking Judder in Foveated Rendering with Phase-Shifted Noise}

\begin{document}

\teaser{
 \includegraphics[width=\linewidth]{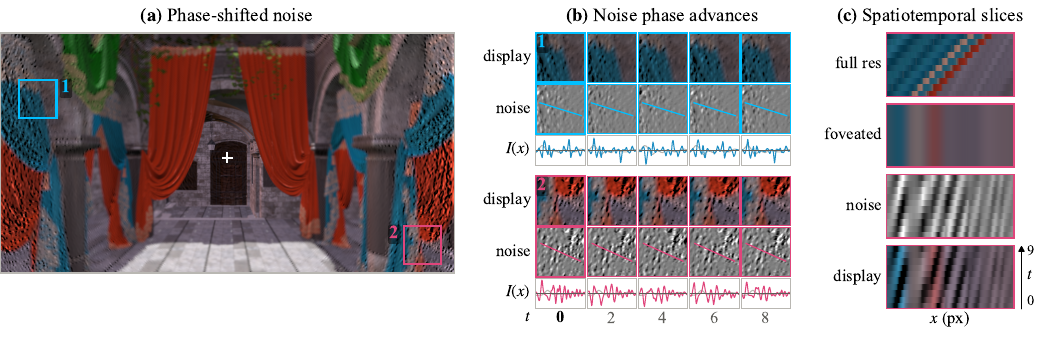}
 \centering
\caption{Rendering at low frame rates saves computation time but introduces judder.
We reduce the visibility of peripheral judder in foveated rendering by adding procedural noise and advancing the phase of the noise on each frame, while the underlying foveated content is rendered at the lower frame rate.
(a) Foveated frame with phase-shifted noise composited over it, designed to be viewed with a 60 degree total field of view.
(b) Insets from (a) over frames 0 - 8.
Noise in this diagram is amplified to highlight changes.
$I(x)$ shows the intensity along the lines for the noise (coloured lines) and the foveated content (gray lines).
(c) Spatiotemporal slice along the pink line with our rendering method.
The full spatiotemporal resolution image (top) is the ground-truth image.
}
    \label{fig:teaser}
}

\maketitle
\begin{abstract}

Low framerates during rendering arise when frame time exceeds the rendering budget, or can be introduced deliberately to save computation, for example in foveated rendering, where the periphery can be rendered at a reduced spatial and temporal resolution.
However, low frame rates also cause non-smooth motion, or judder, degrading perceived image quality.
Judder manifests as aliasing in the spatiotemporal frequency domain; temporal filtering, such as blurring, can suppress it, but at an additional computational cost and incur a larger loss of detail.
Prior work has shown that inexpensive synthesized noise can compensate for the loss of spatial detail and velocity perception caused by foveation, but its potential for restoring motion cues lost to reduced temporal resolution has not yet been explored.
We show that such noise can itself restore these cues: translating a Gabor kernel reduces to a phase shift, and modulating a kernel phase creates a perception of smooth motion without having to fully rerender the noise.
We derive a method that, given the on-screen angular velocity and the temporal rendering rate, selects a spatial frequency noise band such that phase shifts perceptually mask judder without introducing motion aliasing of their own.
User experiments show that our technique substantially expands the range of angular velocities over which motion at reduced temporal resolution remains smooth, without objectionable degradation of image quality.


\end{abstract}
\section{Introduction}

Interactive computer graphics incurs high computational cost to generate high quality graphics.
As spatiotemporal display resolutions continue to grow, especially for Virtual Reality (VR) and Augmented Reality (AR) hardware, shading costs increase drastically.
More recently, such displays have commonly been powered by mobile devices with vastly reduced rendering and power budgets.
Foveated rendering \cite{guenter2012foveated, vaidyanathan2014coarse, patney2016towards} exploits reduced visual acuity in peripheral vision by reducing sample density in the periphery \cite{stengel2016adaptive, polychronakis2021emulating}, vastly reducing shading costs.
Reducing sample density, however, removes frequencies that remain detectable though not resolvable \cite{thibos1987vision}; reducing too much degrades perceived image quality, while achieving a larger performance gain.

Foveated rendering produces \textit{peripheral metamers} \cite{freeman2011metamers}, images that appear the same in peripheral vision, yet differ in content, and recent methods have synthesized them more cheaply than rendering the ground truth image \cite{walton2021beyond}.
Noise synthesis with Gabor noise has pushed spatial resolution reduction even further and maintained high image quality \cite{tariq2022noise}, while \cite{tariq2024towards} noted that spatial loss from foveated rendering also removes spatiotemporal energy, worsening velocity perception, and restored it with a motion metamer.
However, these methods render every frame at the full temporal resolution of the display.
Foveation could equally reduce the rendering rate in the periphery, saving further computation, but rendering below the display rate can manifest as motion judder, or noticeably discretized motion, which becomes more objectionable at higher velocities \cite{chapiro2019luminance, denes2020perceptual}.
In frequency terms, judder is temporal aliasing: moving content sampled at a reduced rate produces copies of its spectrum, whose magnitude and location depend on the sampling rate and velocity; judder becomes visible when these copies fall within the window of visibility \cite{watson1986window}.
Temporal filtering such as motion blur \cite{korein1983temporal}, and spatial low-pass filtering such as foveation, can attenuate this content, but at an even greater cost of spatial detail.
Temporal antialiasing (TAA) \cite{yang2020survey} addresses spatial aliasing, yet will not supply motion cues due to reduced temporal resolution.
Additionally, advancing already synthesized noise on display frames for which no new content is rendered incurs substantially lower cost than rendering those frames directly or filtering them, and this advantage becomes more pronounced in complex scenes.
Foveated metamer pipelines already synthesize noise to restore a removed band, improving apparent resolution \cite{tariq2022noise} or velocity perception \cite{tariq2024towards}.
We now investigate whether noise synthesis can also compensate for reduced \textit{temporal} resolution, asking: \textit{can noise synthesis reduce the visibility of judder beyond what filtering alone achieves?}

In this paper, we show that it can: filtering weakens the aliased copies by attenuating the content they replicate, at the cost of detail, and synthesized noise restores that detail while moving at the full display rate.
Notably, the noise's sampling copies are at the full display rate, and do not introduce perceptible judder.
We refer to this reduction in the visibility of judder by the added noise as \textit{masking}.
Building on previous work on noise metamers \cite{lagae2009procedural, tariq2022noise}, we reconstruct orientation-aligned local first and second order statistics from only the foveated image.
We exploit the fact that translating a Gabor kernel is equivalent to a phase shift to advance the noise every display frame without resynthesis, conveying motion between rendered frames consistent with energy-based models of motion perception \cite{adelson1985spatiotemporal} (Section~\ref{sec:noise-synthesis}).
This process happens far more efficiently than recomputing the entire scene content or other temporal filtering methods.
We first perform a perceptual experiment to investigate peripheral motion judder, finding that rendering high temporal frequency noise over low temporal frequency render content improves smoothness perception at much higher velocities than foveation without noise synthesis (Section~\ref{sec:pilot-study}).
We then derive an analytical model that, given the temporal sampling rate, motion flow, and underlying frequency content, selects a noise frequency whose phase shifts convey motion between rendered frames without aliasing themselves (Section~\ref{sec:method}).
In our validation experiment, we show that this method masks judder to a large degree without a significant loss in image quality (Section~\ref{sec:validation-experiment}).

\section{Related Work}
Our work builds upon a large body of literature in foveated rendering and motion perception.

\subsection{Motion Artifacts}
Motion artifacts manifest in computer graphics in several ways: Screen tearing can arise from desynchronized rendering with the monitor's refresh rate, i.e., low framerate or variable frame timing.
While adaptive synchronization can improve gaming performance using variable frame timing (VFT) \cite{watson2019effects}, VFT can still significantly affect smoothness perception due to individual frames appearing on screen for longer times \cite{klein2024variable}.
Motion judder is a visible discretization of what should be perceived as continuous motion, and often occurs from low display refresh rates.
As individual frames become discernible, motion can appear stuttery, or exhibit judder \cite{denes2020perceptual}.
Short exposures can cause strobing.
Motion blur \cite{korein1983temporal, cook1984distributed, sung2002spatial} can hide motion judder and strobing, but can add computational overhead depending on the filter size, and removes already synthesized high frequency details.

Image persistence at lower refresh rates causes hold-type blur that reduces motion clarity and can impact aiming performance \cite{toth5288560monitor, tokey2025pushing, tokey2026effects}.
Displaying the signal for less time can alleviate hold-type blur, but lowering the display time below the Critical Flicker Fusion (CFF) will lead to undesirable flickering \cite{krajancich2021perceptual}.
Recently, hardware-level backlight strobing with adaptive synchronization has been released to reduce hold blur and improve motion clarity \cite{burnes2026gsyncpulsar} at 60Hz or higher.
However, temporal aliasing artifacts such as motion judder affect motion quality far more than hold-type blur \cite{mackin2016visibility}, and current hardware methods do not operate below 60Hz, where judder is most visible \cite{chapiro2019luminance, denes2020perceptual, jindal2021perceptual}.

CaMoJAB \cite{jindal2021perceptual} is a model that predicts motion quality based on refresh rate, shading resolution, persistence, and velocity.
It demonstrated that hold-type blur at high speeds can reduce the visibility of spatial aliasing from low shading rates, though this decreases with lower image persistence.

\subsection{Foveated Rendering}
Foveated rendering takes advantage of the reduced visual acuity in peripheral vision \cite{thibos1996characterization} by rendering a reduced sample density as eccentricity increases from a user's gaze fixation.
This has enabled higher levels of compression in early offline use cases \cite{kortum1996implementation, tsumura1996image}.
Real-time applications have seen great performance increases from shading reduction due to the progressively lower resolutions rendered in the periphery \cite{guenter2012foveated, vaidyanathan2014coarse}, with further improvements via contrast enhancement \cite{patney2016towards}.
Similar sample density reductions drastically cut cost in ray and path tracing \cite{polychronakis2021emulating, weier2016foveated, swafford2016user}, rasterization \cite{cantory2025enhancing, stengel2016adaptive, meng2018kernel}, split rendering \cite{cantory2023image}, and level-of-detail schemes for traditional and splatting-based systems \cite{ohshima1996gaze, surace2023gaze, franke2025vr}.

The reduced resolution rendering of foveation can be modeled with a spatially varying Gaussian blur \cite{patney2016towards, tursun2019luminance}.
Our method uses a Gaussian blur filter with a standard deviation, $\sigma$, tuned per scene, as foveation levels have been shown to vary based on scene content.
Then, the only input that our noise synthesis and subsequent phase shifts require to suppress the visibility of judder is the foveated image.

\subsection{Metamerism}
Two images are metamers if their content differs, yet they are indistinguishable from eachother.
Peripheral metamerism is particularly desirable for foveated rendering because it can allow for cheaper rendering, where only image statistics (typically first and second order moments) are matched in a local pixel region rather than true pixel colour at each location \cite{freeman2011metamers}.
Deep learning methods can efficiently generate metamers, yet still often fail to meet real-time rendering requirements \cite{DezJE19, feather2019metamers}.
Recent work cleverly identified \textit{peripheral pooling} as a significant consideration when constructing image metamers, and introduced the first real-time foveated metamer \cite{walton2021beyond}.
Convolutions with Gabor impulses have been used to construct more efficient metamers \cite{tariq2022noise} and motion metamers to restore lost velocity perception \cite{tariq2024towards}.
Metamers of scene \textit{understanding}, generated via diffusion from peripheral gist and foveal detail, have also been explored recently \cite{raina2026generating}.

However, prior metamer synthesis methods generate new metamers for each foveated frame, and do not attempt to create metamers to estimate motion when a new frame has not yet been rendered; our work efficiently generates metamers, and modulates them in intermediary frames by estimating motion velocity and applying phase shifts to already constructed noise.
We find that this can successfully mask motion judder without incurring a high decrease in perceived image quality.

\subsection{Temporal Reprojection}

Temporal reprojection techniques have been used to avoid rendering an entire new frame from scratch every time the display requires an update, motivated in part by the fact that frame-to-frame, most content remains almost identical.
Coarse deformable grids have been used to morph up to two frames from a previously rendered frame \cite{didyk2010perceptually}.
Artifacts from reprojection are mitigated by blurring the morph, and then resharpening.
In VR, asynchronous reprojection has been used to avoid missing frames entirely.
Asynchronous Timewarp (ATW) \cite{van2016asynchronous, oculusvr} warps the most recently rendered frame to a user's latest head pose update immediately before display.
As it only handles head rotations,  it does not account for camera translation or object motion, missing disocclusions.
Asynchronous Spacewarp (ASW) \cite{beeler2016asynchronous} extrapolates every other frame entirely, utilizing motion vectors to handle movement.
However, artifacts that occur from this grow as the number of frames reprojected increases.
Our method aims to mask judder perception for very little render cost compared to other temporal reprojection methods, and is complementary to reprojection.

\begin{figure}[tbp]
    \centering
    \begin{subfigure}{0.19\columnwidth}
        \includegraphics[width=\linewidth]{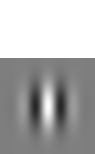}
        \caption{}
        \label{fig:gabor-a}
    \end{subfigure}
    \begin{subfigure}{0.19\columnwidth}
        \includegraphics[width=\linewidth]{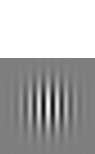}
        \caption{}
        \label{fig:gabor-b}
    \end{subfigure}
    \begin{subfigure}{0.19\columnwidth}
        \includegraphics[width=\linewidth]{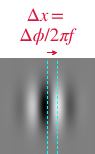}
        \caption{}
        \label{fig:gabor-c}
    \end{subfigure}
    \begin{subfigure}{0.19\columnwidth}
        \includegraphics[width=\linewidth]{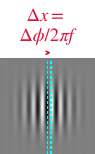}
        \caption{}
        \label{fig:gabor-d}
    \end{subfigure}
    \begin{subfigure}{0.19\columnwidth}
        \includegraphics[width=\linewidth]{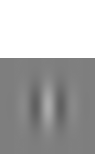}
        \caption{}
        \label{fig:gabor-e}
    \end{subfigure}
    \caption{
    A Gabor at a low \textbf{(a)} and high \textbf{(b)} spatial frequency $f$, sharing the same Gaussian envelope.
    \textbf{(c, d)} The same phase step $\Delta\phi$ applied to each: the envelope stays fixed while the sinusoid translates by $\Delta x = \frac{\Delta \phi}{2\pi \, f}$ (dashed markers).
    The same phase step yields lower displacement at the higher frequencies, requiring lower frequency Gabors to mask $\rho_{J}$.
    \textbf{(e)} Reducing the amplitude $A$ lowers the contrast of the same pattern.
    }
    \label{fig:gabor-panels}
\end{figure}

\section{Noise Synthesis}
\label{sec:noise-synthesis}
We render the base noise with the method of \cite{tariq2022noise}, which uses only the foveated frame to synthesize high frequency peripheral noise that is detectable but not resolvable in peripheral vision; by matching local image statistics, it reincorporates texture energy without creating new visual structures.
Gabor kernels (a sinusoid modulated by a Gaussian envelope) \cite{lagae2009procedural} are convolved with sparse random impulses.
The screen is partitioned into a grid of cells, each of which deterministically seeds its impulse locations $\mathbf{c}_i$, a pixel $\mathbf{p}$, then accumulates the truncated sum over nearby impulses to compute the final noise value $N(p)$:
\begin{equation}
N(p) = \sum_{i \in I} A_{i}  e^{-\pi a_{i}^{2} ||p - c_{i}||^{2}} \cos \big( 2\pi f_{i} \langle p - c_{i}, \theta_{i} \rangle + \phi_{i} \big)
\label{eq:gabor-sum}
\end{equation}
where $I$ is the set of impulses, $a_{i}$ is size of the Gaussian envelope, and each kernel is fully described by four parameters: spatial frequency $f_i$, orientation $\theta_{i}$, amplitude $A_i$, and phase $\phi_i$.
Unless noted, spatial frequencies are expressed in cycles per pixel (cpp).

\textit{Frequency}.
Between the peripheral resolution and detection limits, a stimulus registers as contrast energy but cannot be resolved \cite{thibos1987vision}; this is the band the synthesis restores.
Synthesized frequencies are constrained to the Thibos limits, $[F_{L}, F_{H}]$, computed from the resolution and detection limits of \cite{thibos1996characterization} (linearly interpolated across eccentricity $e$).
Foveation constrains the bands further; modeling the eccentricity-dependent foveation as a Gaussian blur with standard deviation $\sigma(e)$, the frequency response falls to ${\sim}1\%$ at $3\sigma_f$, where $\sigma_f(e) = 1/(2 \pi \sigma(e))$.
Each impulse draws $f_{i}$ from a log-normal distribution centered on the band's log-midpoint, the center frequency $F_{C} = \sqrt{F_{L} F_{H}}$, clamped to $[F_{L}, F_{H}]$ \cite{de1982spatial}.
We add a band scale $s \in (0, 1]$ that rescales both bounds to $[s F_{L}, s F_{H}]$, giving the representative frequency $f_{cpp} = s F_{C}$, which the display's pixels per degree converts to cycles per degree, $f_{cpd}$.
Section~\ref{sec:method} derives lower scales as  needed to restore motion cues in juddery scenarios.

\textit{Orientation}.
\label{sec:synth-orientation}
Peripheral vision remains sensitive to the dominant orientation of unresolvable detail \cite{walton2021beyond}.
In static scenes, orienting $\theta_{i}$ such that it follows the local structure of the foveated image, estimated with Sobel filters, is an ideal strategy.
However, to maximize the motion from phase shifting, Gabor orientation can be aligned with motion flow.
As this structure comes from the surviving low frequencies, the orientation is spatiotemporally stable.

\textit{Amplitude}.
\label{sec:synth-amplitude}
The reconstructed amplitude should reproduce the contrast of the detail that is missing due to foveation.
Following \cite{tariq2022noise}, we infer it from the surviving band just below the foveation cut-off, justified by the power-law spectrum of natural images \cite{ruderman1994statistics}.
We decompose the foveated luminance into a Laplacian pyramid $L$, and select, per pixel, the level $l$ whose passband is centered nearest the local cutoff frequency (level $l$'s center frequency is $2^{-(l + 0.5)}$ cpp), and set $A_i = s_k | L_{l}(c_{i}) |$, where $s_k$ is a global gain (Figure~\ref{fig:gabor-e}).
The noise is synthesized only from the luminance channel, so it must be added to the input luminance with contrast enhancement, $Y' = \operatorname{clip}(Y + N)$ \cite{patney2016towards}, and we uniformly rescale the pixel's colour by the ratio $\frac{Y'}{Y}$.
To prevent clipping from distorting the synthesized spectrum, the amplitude is additionally attenuated against local minimum and maximum image bounds before synthesis.

\textit{Phase}.
$\phi_{i}$ (Figures~\ref{fig:gabor-c},~\ref{fig:gabor-d}) is the only parameter we can modulate to advance a Gabor kernel.
As we will explore in Section~\ref{sec:method}, we can advance the phase at the full display rate to convey motion.

\section{Exploratory Perceptual Study}
\label{sec:pilot-study}

\begin{figure}[tbp]
    \centering
    \begin{subfigure}{0.48\linewidth}
        \includegraphics[width=\linewidth]{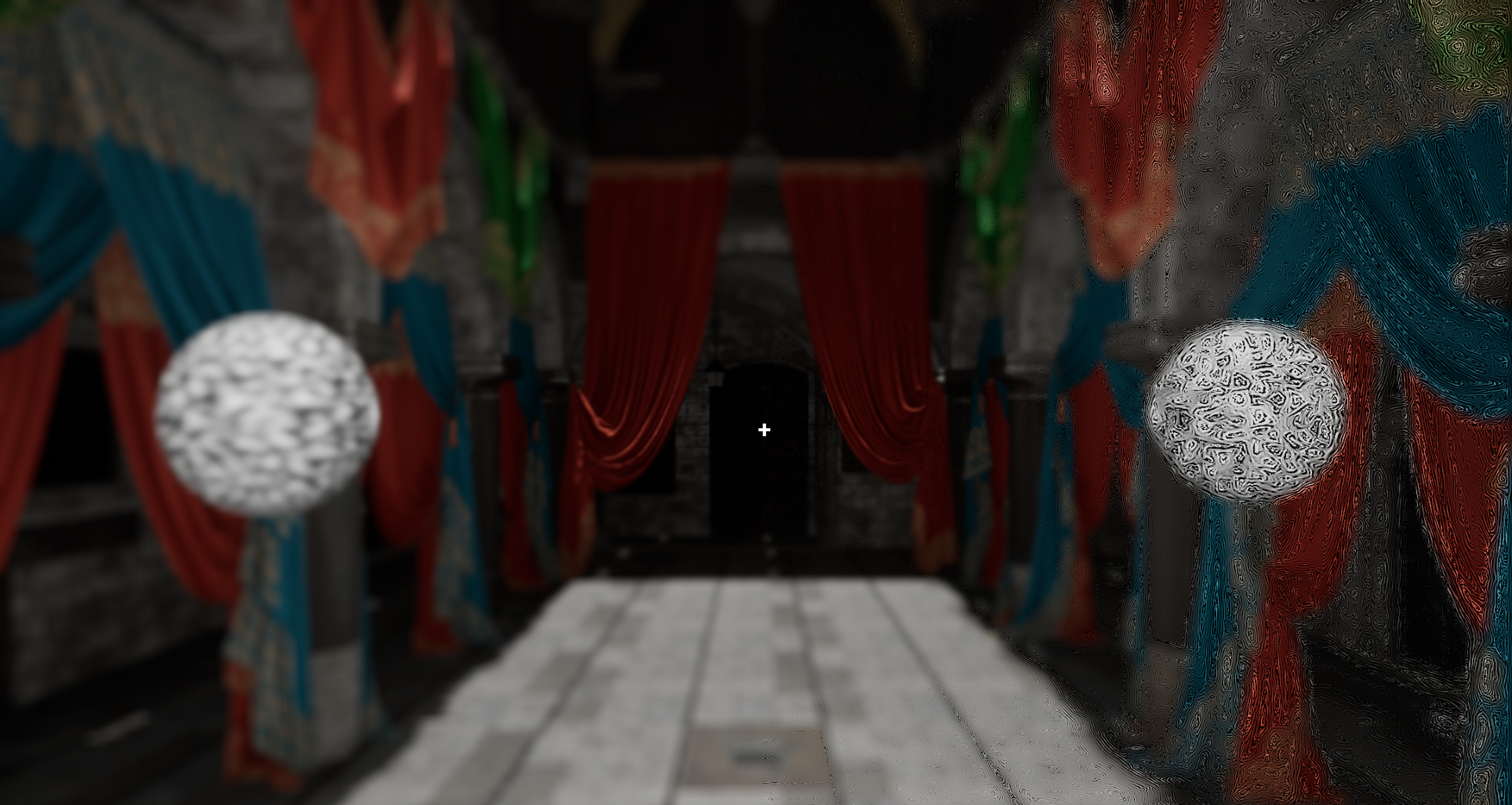}
        \caption{}
        \label{fig:pilot-study-stimulus}
    \end{subfigure}\hfill
    \begin{subfigure}{0.48\linewidth}
        \includegraphics[width=\linewidth]{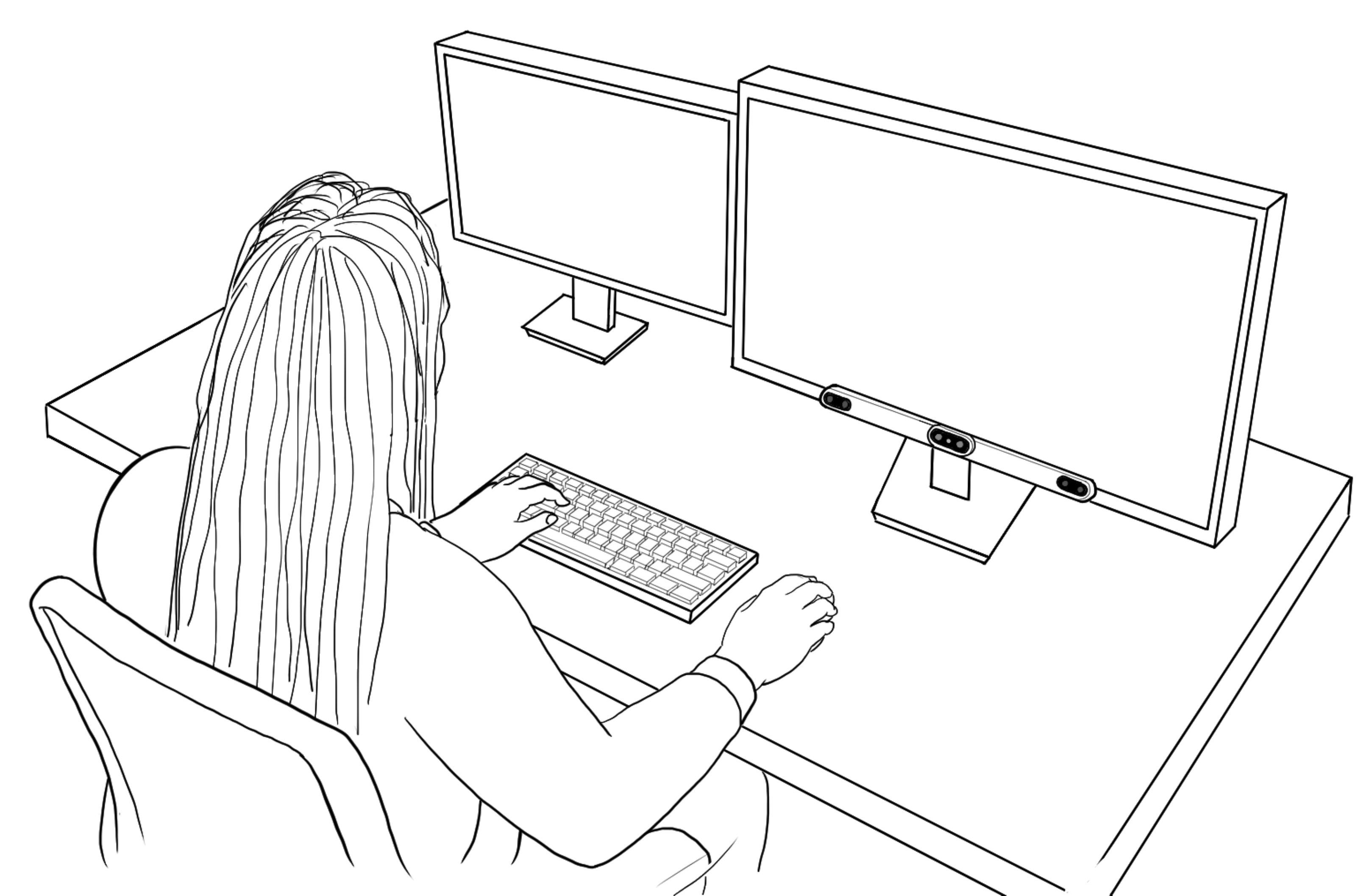}
        \caption{}
        \label{fig:pilot-study-setup}
    \end{subfigure}
    \caption{
    \textbf{(a)} Left: Example of a Perlin noise textured sphere under Gaussian blur.
    Right: The same sphere with noise rendered at a lower frequency scale, $s = 0.5$.
    In the exploratory study, noise was applied to the entire test side of the screen at a single uniform frequency scale; in the validation study, the scale follows our full model in Eq.~\ref{eq:sstar}.
    In the study, the center 10 degree radius of vision was grayed out to focus on peripheral vision (not pictured).
    \textbf{(b)} Experimental setup for validation experiment (Section~\ref{sec:validation-experiment}).
    }
    \label{fig:pilot-study}
\end{figure}

\begin{figure}[tbp]
    \centering
    \includegraphics[width=\linewidth]{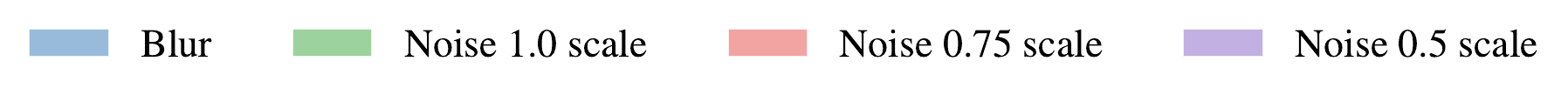}\\[3pt]
    \begin{subfigure}{0.49\linewidth}
        \includegraphics[width=\linewidth]{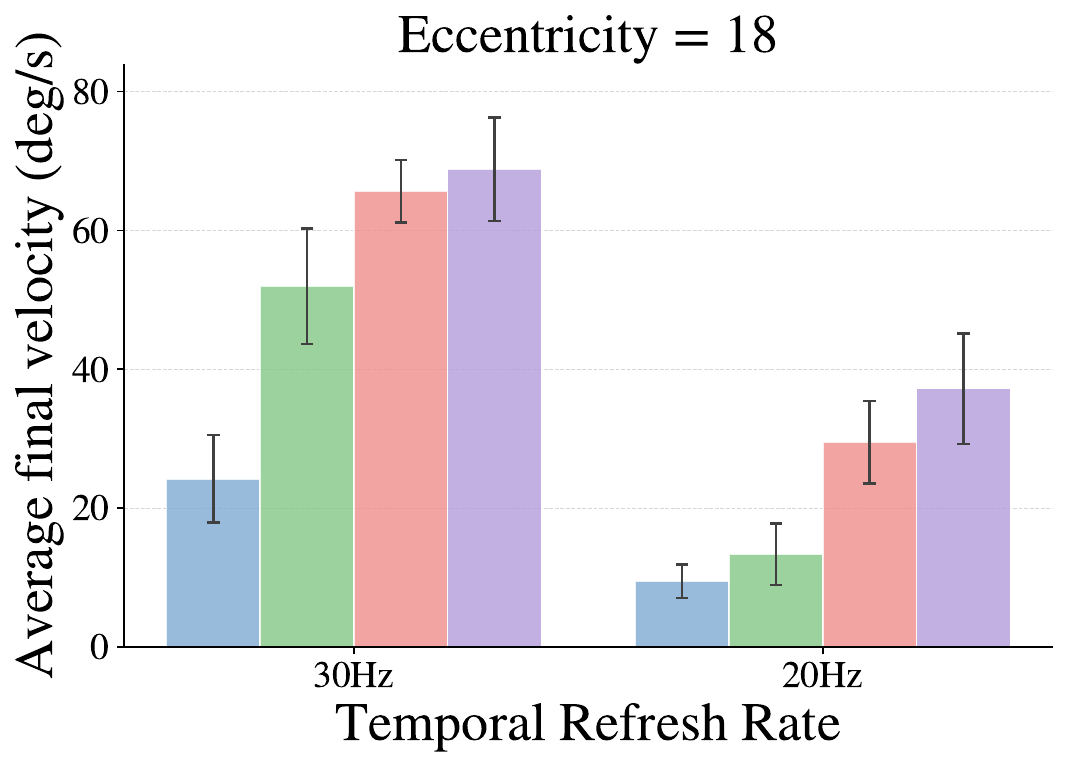}
        \caption{}
        \label{fig:exp0-ecc18}
    \end{subfigure}\hfill
    \begin{subfigure}{0.49\linewidth}
        \includegraphics[width=\linewidth]{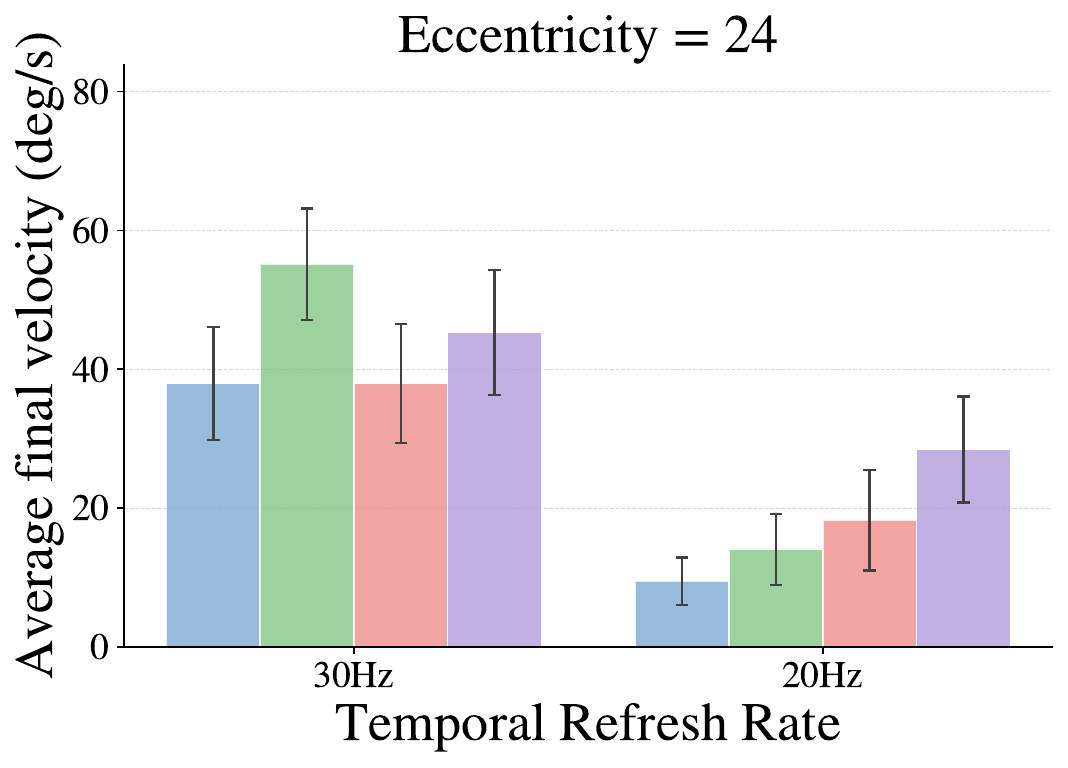}
        \caption{}
        \label{fig:exp0-ecc24}
    \end{subfigure}
    \\[4pt]
    \begin{subfigure}{0.49\linewidth}
        \includegraphics[width=\linewidth]{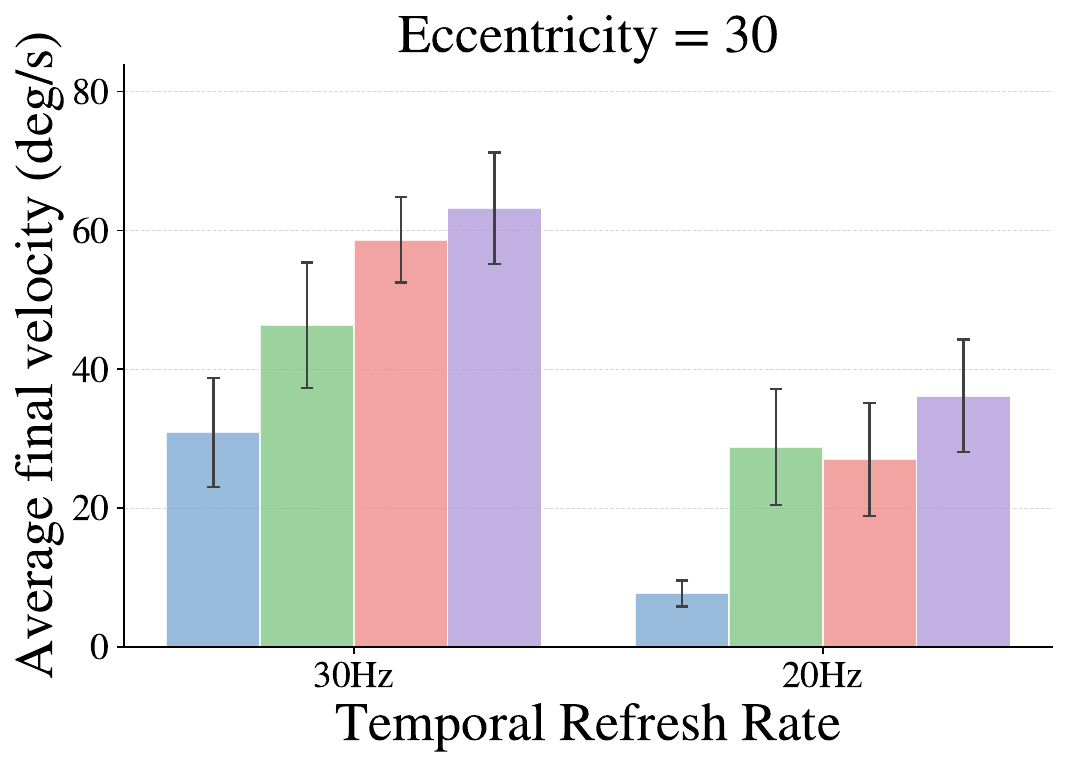}
        \caption{}
        \label{fig:exp0-ecc30}
    \end{subfigure}\hfill
    \begin{subfigure}{0.49\linewidth}
        \includegraphics[width=\linewidth]{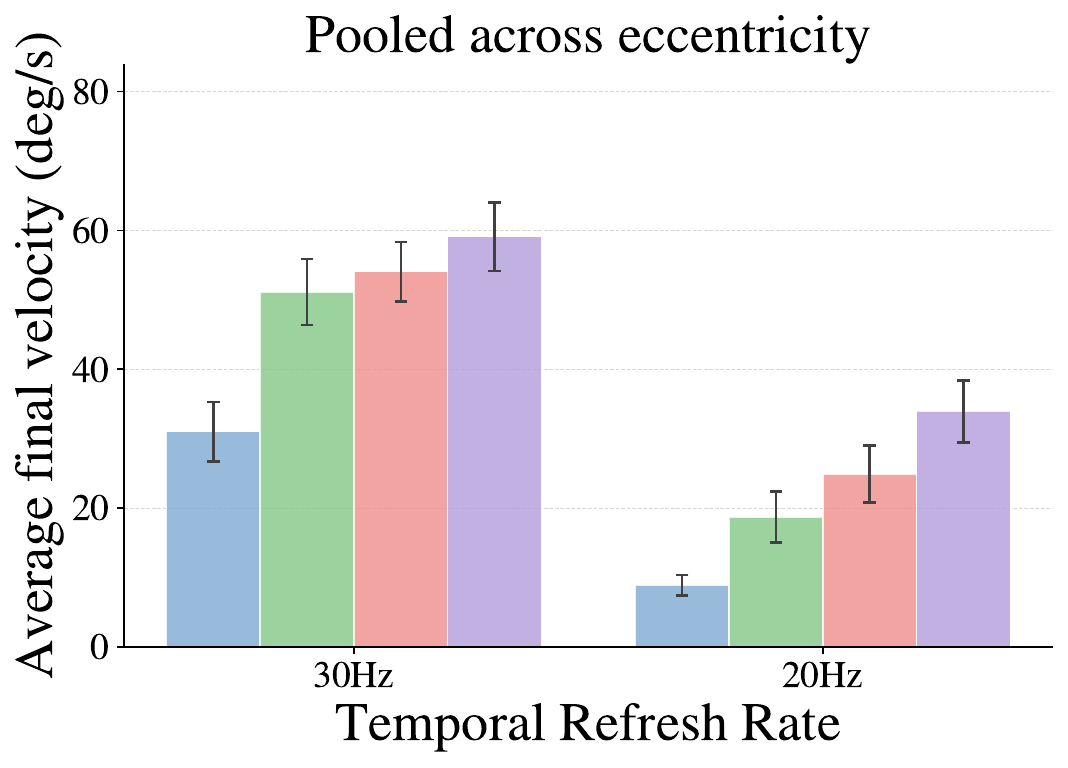}
        \caption{}
        \label{fig:exp0-pooled}
    \end{subfigure}
    \caption{
    Matched velocities at eccentricity \textbf{(a)} $18^{\circ}$, \textbf{(b)} $24^{\circ}$, \textbf{(c)} $30^{\circ}$, and \textbf{(d)} pooled across eccentricities.
    Error bars are standard error of the mean.
    Generally, as the noise scale decreased, higher matched velocities monotonically increased.
    }
    \label{fig:exp0-velocities}
\end{figure}

In spatiotemporal energy models \cite{adelson1985spatiotemporal}, motion perception pools the motion energy extracted by oriented filters across frequency bands.
With a reduced temporal resolution, the surviving low frequency bands carry temporally aliased motion energy, whereas the synthesized noise band, though estimated from the foveated content, can move at the full display rate; if motion perception indeed pools these signals, this coherent motion may perceptually mask the aliased copies.
Noise in the detectable but not resolvable band uses a scale $s = 1$ (Section~\ref{sec:noise-synthesis}).
However, because the same phase step translates a lower frequency Gabor further (Figures~\ref{fig:gabor-c}, \ref{fig:gabor-d}), lower frequencies can convey greater displacement between frames, and therefore represent faster motion.
We therefore ask: \textit{can noise updated at the display rate over content rendered at a lower rate mask judder, and does lowering its spatial frequency extend masking to higher velocities?}
We ran a small IRB approved pilot study (n=5), scaling the noise of Section~\ref{sec:noise-synthesis} to lower frequencies with a constant scale factor $s$.
In this study, we fix the frequency scale within a trial, and the noise is moved via reprojection following motion flow while the underlying, low frequency image content remains static during intermediary frames.

\textit{Setup.} Stimuli were shown on a Dell 2725QC 27 inch 4K 120Hz monitor with peak luminance set to $150.0\,cd/m^{2}$, viewed from 42.8cm, on a machine with an Intel 12900K and an NVIDIA RTX 4070 Ti Super.
Participants were eye tracked with a Tobii Eye Tracker 5.
All rendering was restricted to 60Hz to prevent screen tearing or unintentional judder.
Foveation was simulated with a Gaussian blur that increases with eccentricity, using a very high blur value ($\sigma = 0.34$) based on prior work \cite{tursun2019luminance, tariq2022noise}.
Five participants, all of whom have worked or published in computer graphics, took part (ages 28-30, 3 male, 2 female).

\textit{Stimulus.}
Textured spheres of $3$ degree angular radius, centered at eccentricity $e = 18, 24,$ or $30$ degrees and placed symmetrically about the fixation center, translated back and forth vertically in the Sponza scene \cite{McGuire2017Data, intel2022samples}.
Textures were a high frequency grass texture and a low frequency Perlin Noise \cite{perlin1985image} texture (Figure~\ref{fig:pilot-study}).
The test side was rendered either with the radially increasing Gaussian blur alone (foveation), or with noise at one of three scales: high ($s = 1$), medium ($s = 0.75$), and low ($s = 0.5$).
The test stimulus was presented at 60, 30, or 20Hz, with three repeats per condition; 60Hz trials served as catch trials and as an exclusion criterion, since their aliasing copy lies outside the boundary of resolvability (Figure~\ref{fig:judder-boundary}), so users were not expected to adjust the velocity meaningfully.
To focus solely on peripheral judder, the center $10$ degree radius of vision was grayed out.

\textit{Task.} Users fixated on a central cross and were shown two moving stimuli; whenever gaze deviated from the cross, the screen turned black until it returned.
The left side showed a foveated reference at 60 frames per second \cite{tursun2019luminance}; foveation was used as the reference because we expected judder to be less perceptible under Gaussian blur than at full resolution.
Using the method of adjustment, each trial started at approximately $86$ degrees per second, well into the range where pilot testing showed visible judder, and participants adjusted the velocity with the arrow keys; they were instructed to adjust the object velocity until both sides appear equally smooth; noise was applied across the entire test side.
Higher matched velocities indicate stronger reductions in the visibility of judder; a representative image of the setup is shown in Figure~\ref{fig:pilot-study}.

\textit{Results.}
At every rendering rate, noise conditions appeared smooth up to higher velocities than the blur baseline (Figure~\ref{fig:exp0-velocities}).
Users rarely adjusted the velocity in the 60Hz catch trials, as expected for motion without visible judder, so no participants were excluded.
Pooled over eccentricities and textures, matched velocity rose from $31.0$ deg/s under blur to $51.1$-$59.1$ deg/s across noise scales at 30Hz, and from $8.9$ to $18.7$-$33.9$ deg/s at 20Hz, a $1.65$-$3.8\times$ increase in the velocity tolerated before judder is perceptible; we noticed no difference between stimulus textures.
Matched velocity generally increased monotonically as $s$ decreased, but only partially at $e = 24$ degrees and 30Hz, possibly due to the band-pass shape of peripheral contrast sensitivity, whose peak spatial frequency shifts with eccentricity \cite{mantiuk2022stelacsf, krajancich2021perceptual}: lowering the noise frequency can move it towards the sensitivity peak rather than away from it.
These results indicate that even when the underlying low spatial frequency content remains static between rendered frames, intermediary frames can convey motion by rendering synthesized noise along the path of motion.
Lowering the noise frequency clearly allowed smooth motion at higher velocities, but the study does not tell us how far to lower it for a given motion, or how to compute that scale.
We use these insights to derive a velocity-adaptive frequency scale (Section~\ref{sec:method}) and to design our validation experiment, which conveys motion via phase shifts (the one component of noise that can be updated without resampling the noise).

\section{Phase Shifts Mask Motion Judder}
\label{sec:method}
Phase carries the perceptually dominant structure of images: swapping phase spectra between images transfers their recognizable content while amplitude swaps do not \cite{oppenheim1981importance, piotrowski1982demonstration}, and degrading phase destroys the classification of natural images \cite{wichmann2006phase}.
Advancing the phase of each kernel conveys motion at the cost of only a single scalar update per kernel (Section~\ref{subsec:implementation}).
As Figures~\ref{fig:gabor-c},~\ref{fig:gabor-d} show, this displacement scales as noise frequency decreases, so restoring the motion energy removed by low temporal sampling rates requires adapting the noise frequency to the motion.
Our exploratory study (Section~\ref{sec:pilot-study}) confirmed that lower noise frequencies extend masking to higher velocities, but only tested a single fixed scale regardless of motion.
Scenes with complex movement require variable frequency scaling throughout, such that the noise is at an appropriate frequency to convey motion, but not so low as to become distracting or resolvable.
To choose the noise frequency, we first model judder as temporal aliasing, which identifies the aliased frequencies whose visibility the noise must reduce (Section~\ref{sec:judder}).
We use a contrast sensitivity model to predict when judder and the noise are visible (Section~\ref{sec:visibility}).
Applying the same aliasing model to the phase-shifted noise then yields a velocity-adaptive frequency scale $s^{*}$ for the synthesized motion signal added to the content (Section~\ref{sec:phase-motion}), whose perceptual effect we measure in a user study (Section~\ref{sec:validation-experiment}).

\subsection{Motion Judder}
\label{sec:judder}

\begin{figure}[tbp]
    \centering
    \begin{subfigure}{0.5\linewidth}
        \includegraphics[width=\linewidth]{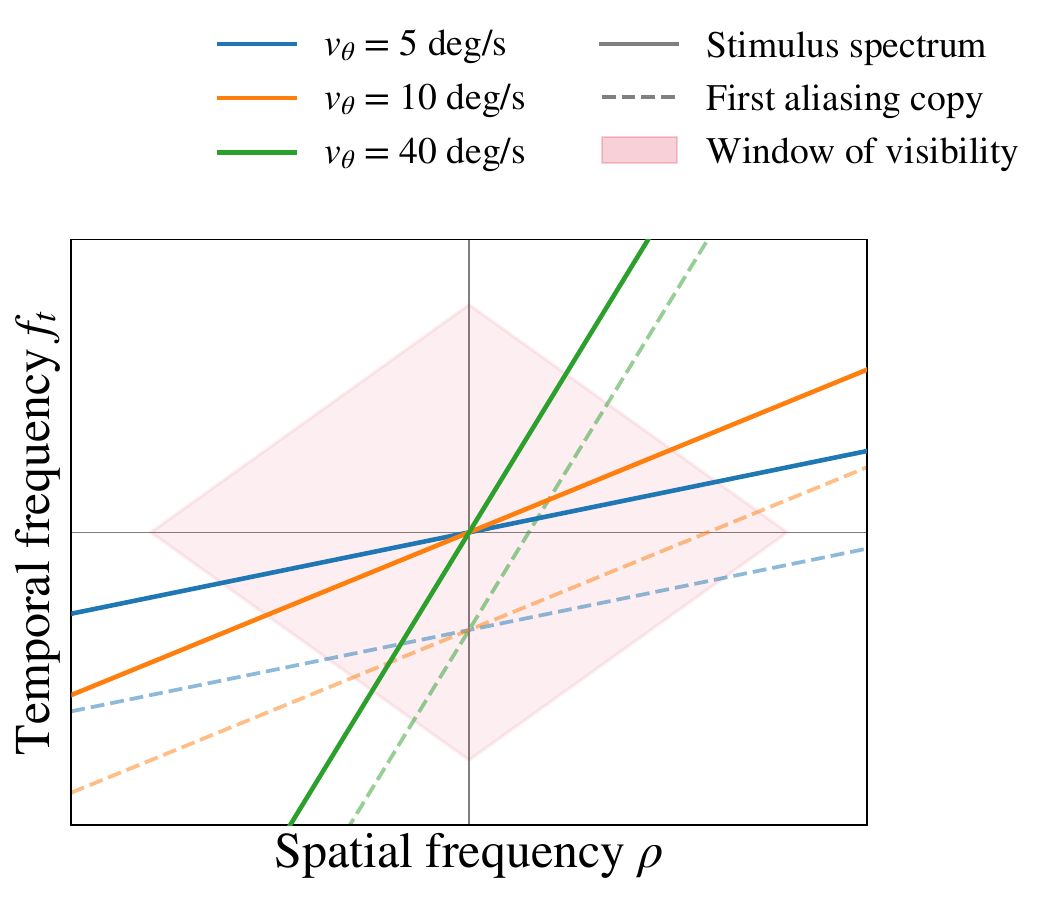}
        \caption{}
        \label{fig:aliasing-a}
    \end{subfigure}\hfill
    \begin{subfigure}{0.5\linewidth}
        \includegraphics[width=\linewidth]{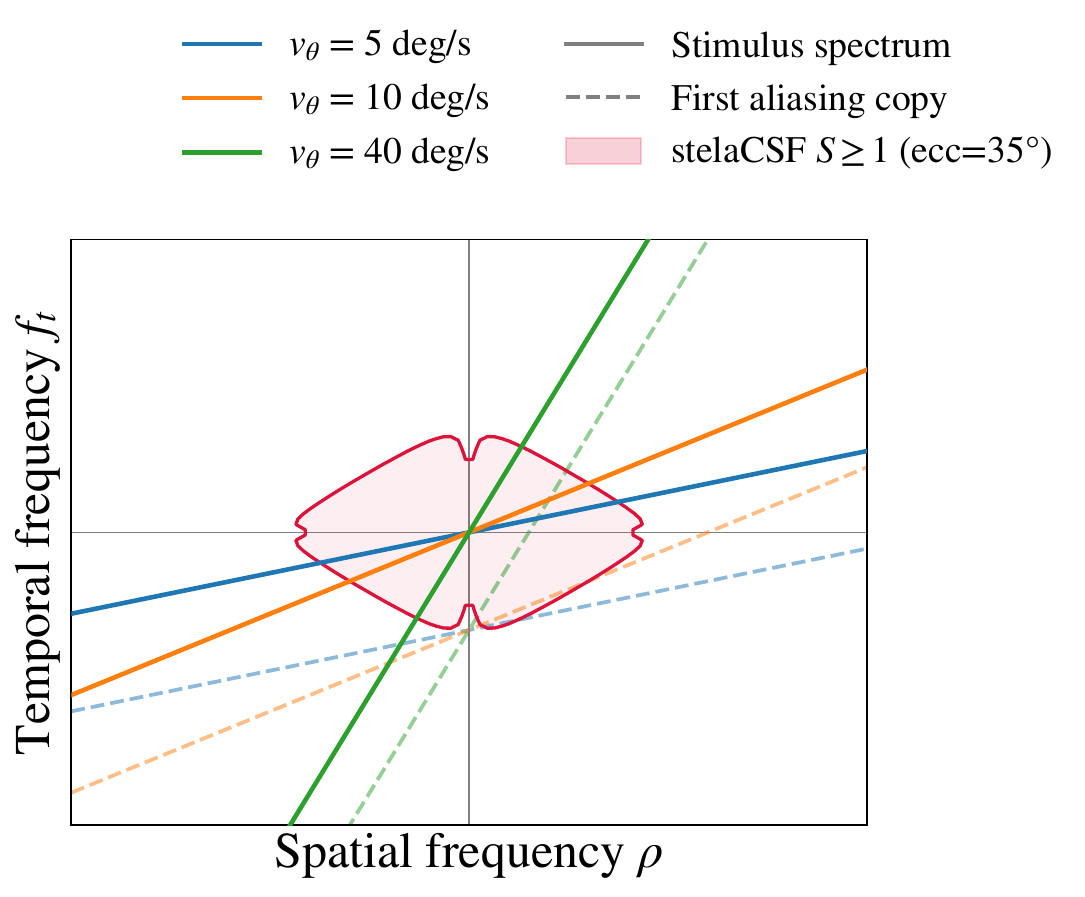}
        \caption{}
        \label{fig:aliasing-b}
    \end{subfigure}
  \caption{
    \textbf{(a, b)} The boundary of visibility of Watson's Window of Visibility (a) and StelaCSF (b), and the aliasing (dashed lines) of signals (solid lines through origin).
    Higher angular velocities pull the aliasing copies further into the boundary of visibility towards the origin.
    Outside the boundary of visibility ($S < 1$ in StelaCSF), the aliasing copies are not visible.
    }
    \label{fig:aliasing}
\end{figure}

Consider content with spatial frequency $\rho$ translating at angular velocity $v_\theta$ (degrees per second).
The translation gives a temporal frequency:
\begin{equation}
    f_t = \rho \,  v_{\theta}
    \label{eq:motion-line}
\end{equation}
In the spatiotemporal frequency domain, $(\rho, f_{t})$ spans a line through the origin with slope $v_{\theta}$.
Sampling this motion at the discrete rate $R$ replicates its spectrum at temporal offsets of $\pm k R$, $k \in \mathbb{N}$.
When these aliasing copies fall within the window of visibility \cite{watson1986window}, motion judder becomes visible.

The first aliasing copy is the most egregious, as it lies at the lowest and therefore most visible frequencies \cite{denes2020perceptual}: If it has temporal frequency $R$, its spatial frequency is located at $\rho_{J}$, the spatial frequency of judder:
\begin{equation}
    \rho_{J} = \frac{R}{v_{\theta}}
\label{eq:rho-judder}
\end{equation}
in cycles per degree.
Higher angular velocities decrease the visible frequency, shifting the aliasing copy closer to the original line, and increasing judder visibility.
Judder visibility further scales with how much content energy lies near $\rho_{J}$.

In Figure~\ref{fig:aliasing}, we demonstrate how low sampling rates lead to visible aliasing copies.
At low refresh rates, the original signal (line through the origin) has aliasing copies offset from the origin, which cross through the boundary of visibility.
Higher angular velocities produce higher slopes, rendering alias copies visible at higher refresh rates (green lines in Figures~\ref{fig:aliasing-a} and~\ref{fig:aliasing-b}), while extremely low angular velocities may not be within the window of visibility even for very low temporal sampling rates (blue lines).

Whether judder is visible depends on where the aliased energy near $\rho_{J}$ falls relative to this boundary; as the boundary shifts with eccentricity, and our noise targets peripheral vision, we use a contrast sensitivity model to predict when this energy is visible and whether the noise is.

\subsection{Visibility of Judder and Noise}
\label{sec:visibility}
Contrast Sensitivity Functions (CSFs) \cite{kelly1979motion, krajancich2021perceptual} model visual sensitivity to luminance contrast at varying spatial frequencies, and describe limits of the spatiotemporal resolution of the human visual system.
Influencing factors include eccentricity, luminance, stimulus size (area), orientation, and possibly more, depending on the model used.
We use StelaCSF \cite{mantiuk2022stelacsf} to predict the sensitivity to a stimulus, $S$:
\begin{equation}
    S = S_{csf}(f_{cpd}, R, e, L, A)
    \label{eq:stelacsf}
\end{equation}
where $f_{cpd}$ is the spatial frequency of the stimulus in cycles per degree, $R$ is its temporal frequency in Hz, $e$ is the eccentricity in degrees, $L$ is the adaptation luminance in $cd/m^{2}$, and $A$ is the stimulus area in square degrees.

\begin{figure}[htbp]
    \centering
    \includegraphics[width=\columnwidth]{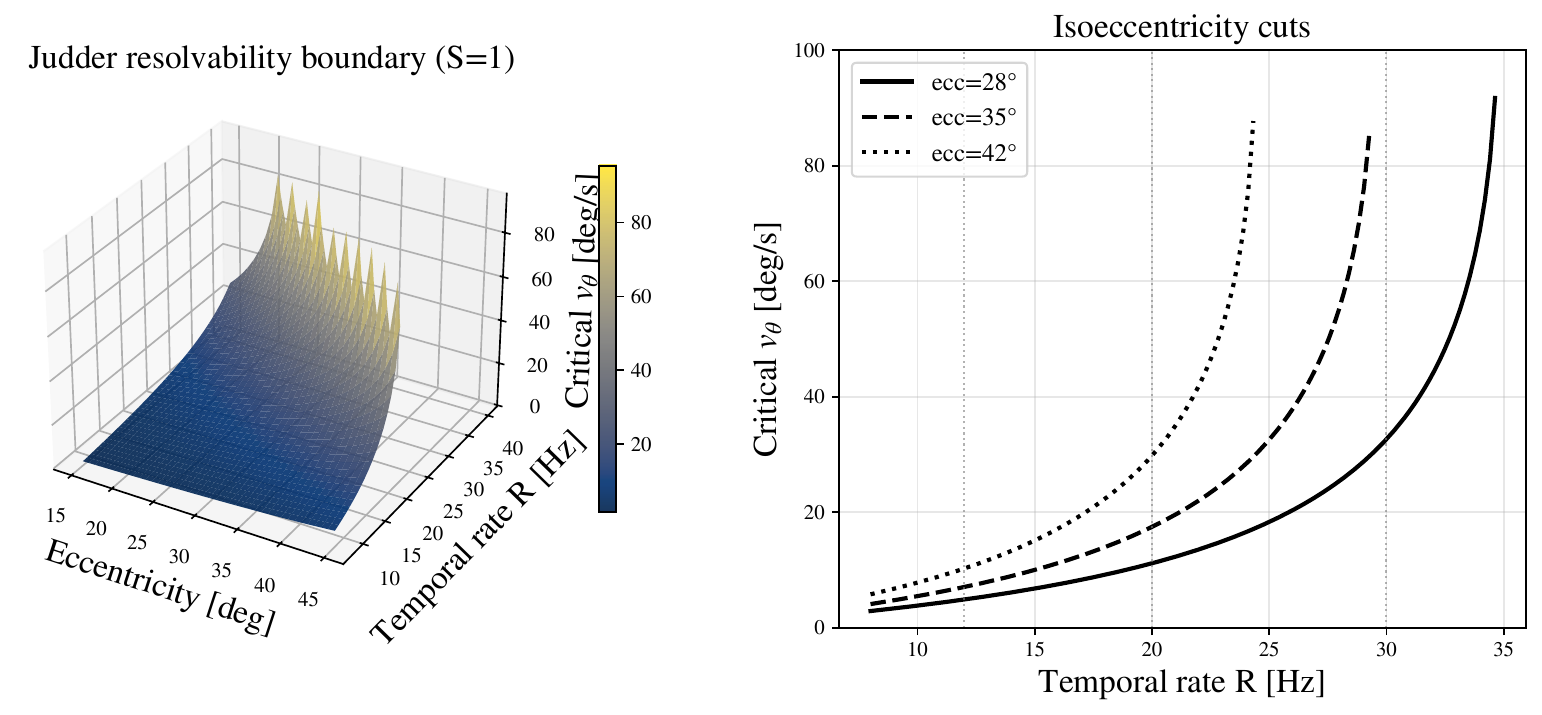}
    \caption{StelaCSF prediction of the boundary of resolvability. Left: The surface predicts the boundary for where judder is resolvable (under the curve is not resolvable, or looks smooth). Right: For a given refresh rate and picked eccentricities, this models the critical $v_{\theta}$, the angular velocity above which judder is resolvable, based on StelaCSF's predictions. }
    \label{fig:judder-boundary}
\end{figure}

\begin{figure*}[tbp]
    \centering
    \begin{subfigure}{0.32\textwidth}
        \includegraphics[width=\linewidth]{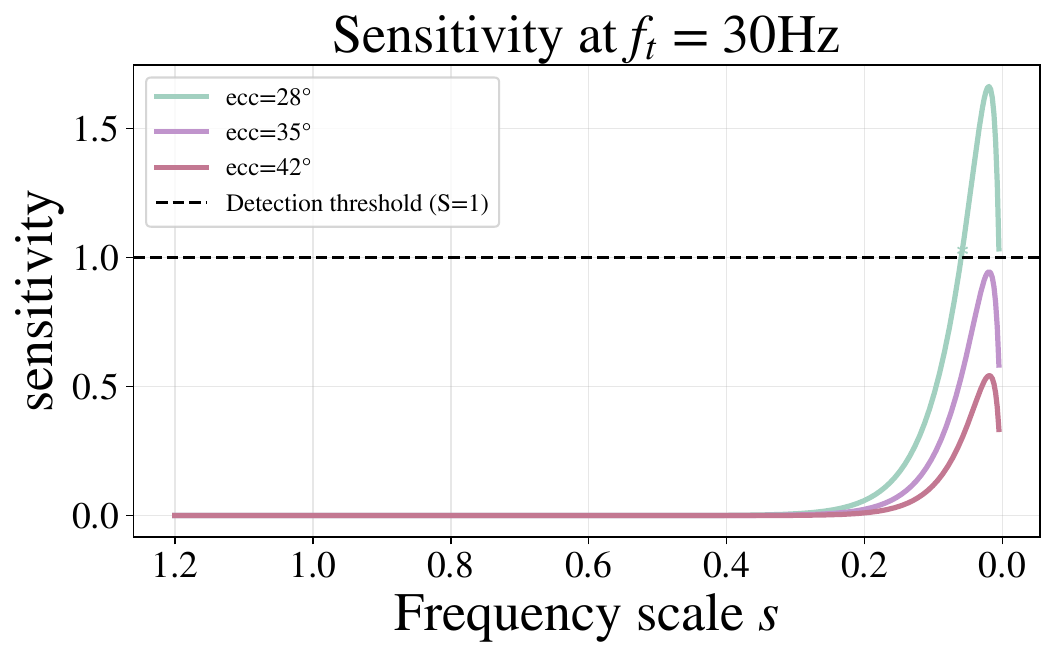}
        \caption{}
        \label{fig:gabor-sens-30}
    \end{subfigure}\hfill
    \begin{subfigure}{0.32\textwidth}
        \includegraphics[width=\linewidth]{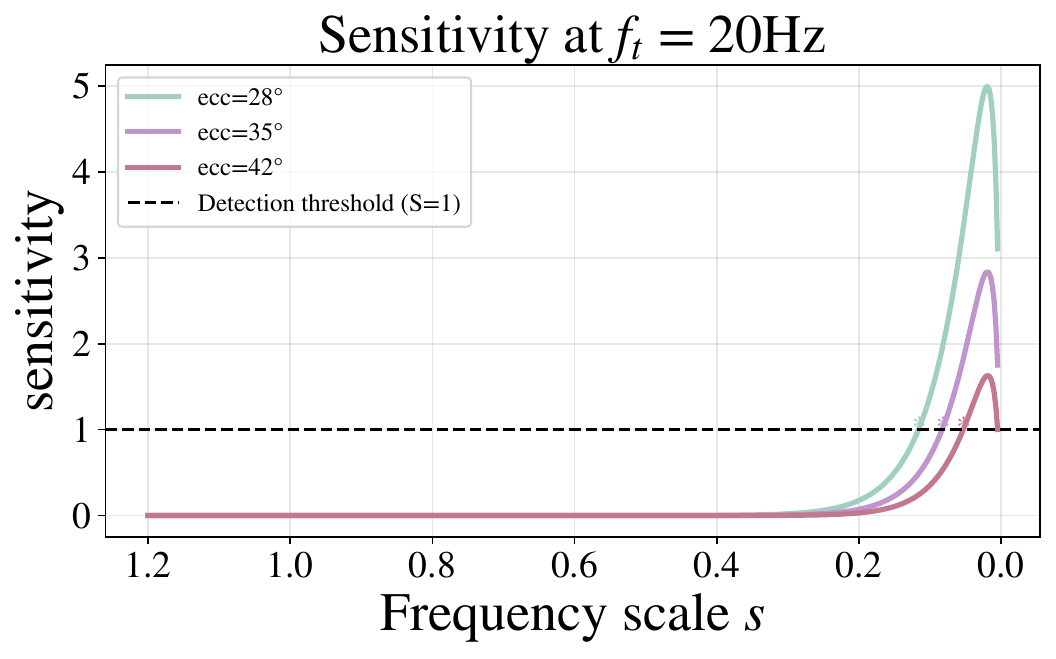}
        \caption{}
        \label{fig:gabor-sens-20}
    \end{subfigure}\hfill
    \begin{subfigure}{0.32\textwidth}
        \includegraphics[width=\linewidth]{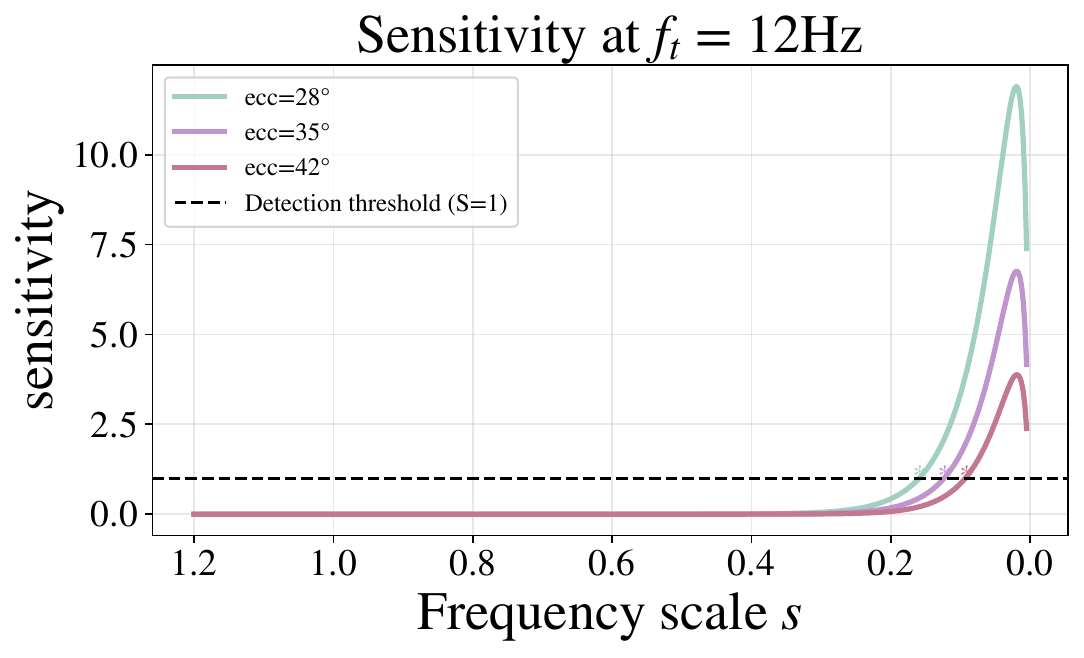}
        \caption{}
        \label{fig:gabor-sens-12}
    \end{subfigure}
    \caption{
    StelaCSF sensitivity to the synthesized noise as a function of the frequency scale $s$, at temporal frequencies \textbf{(a)} $30$Hz, \textbf{(b)} $20$Hz, and \textbf{(c)} $12$Hz.
    Curves are shown for the three tested eccentricities, and the marker on each curve is the scale at which the noise crosses the threshold $S = 1$.
    The stimulus size is modeled as a Gaussian envelope with $\sigma = 1$ degree.
    }
    \label{fig:gabor-sensitivity}
\end{figure*}

We can use StelaCSF as a proxy to predict the visibility of judder.
Content translating at angular velocity $v_\theta$ is sampled at a rate $R$ equal to its own temporal frequency at $\rho_J = \frac{R}{v_{\theta}}$ (Eq.~\ref{eq:rho-judder}), where the first aliasing copy falls to zero temporal frequency.
As a proxy for the visibility of judder, we evaluate Eq.~\ref{eq:stelacsf} at $(\rho_{J}, R)$, and when $S \le 1$, it predicts the sampled motion is below the visibility threshold.
For a fixed content rate and eccentricity, this defines a critical angular velocity above which judder becomes visible, which we plot in Figure~\ref{fig:judder-boundary}.
At the 20Hz and 12Hz content rates we test, these velocities are low, roughly 4 to 30 deg/s across $28$-$42^{\circ}$ eccentricity, making judder visible over much of the peripheral motion in our scenes; this is the range in which the noise must reduce its visibility.
The same model predicts the visibility of the synthesized noise as a function of its frequency scale and temporal frequency (Figure~\ref{fig:gabor-sensitivity}): over the scales we consider, lowering $s$ raises the predicted sensitivity to the noise.

This sets up the central trade-off for our method;
because the same phase step displaces a lower frequency Gabor further (Figures~\ref{fig:gabor-c},~\ref{fig:gabor-d}), lowering the noise frequency lets its phase shifts convey faster motion, yet at the same time, lowering it raises the predicted visibility of the noise (Figure~\ref{fig:gabor-sensitivity}), potentially leading to noise that is visible.
To minimize the visibility of the noise, we therefore seek the highest frequency whose phase shifts can still convey the motion.

\subsection{Phase Conveys Motion}
\label{sec:phase-motion}

\begin{figure}[htbp]
    \centering
    \begin{subfigure}{0.30\linewidth}
        \includegraphics[width=\linewidth]{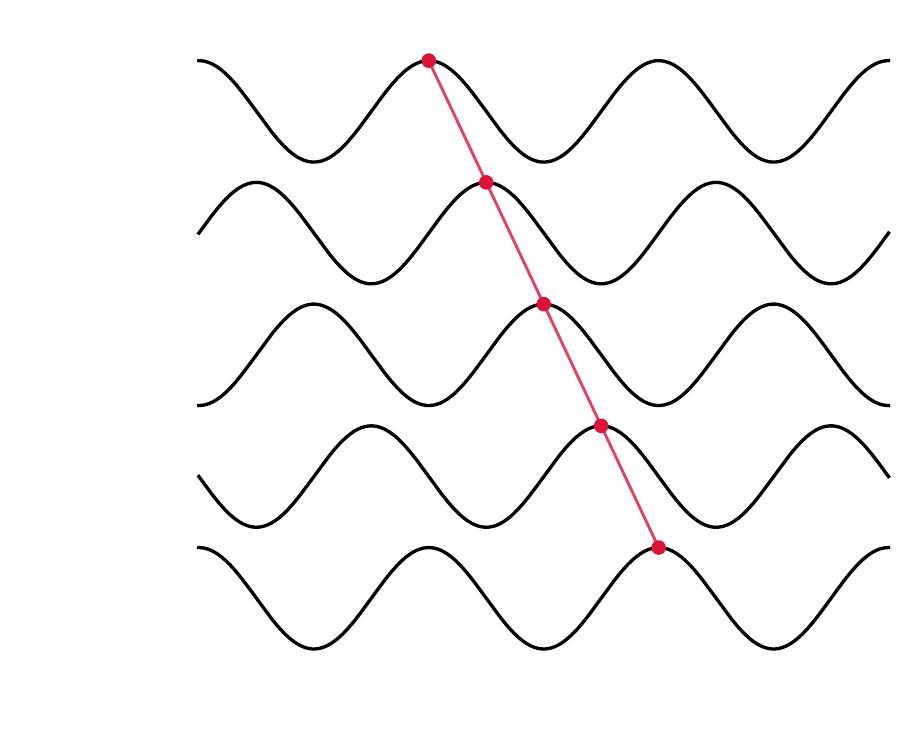}
        \caption{}
        \label{fig:phase-b}
    \end{subfigure}\hfill
    \begin{subfigure}{0.67\linewidth}
        \includegraphics[width=\linewidth]{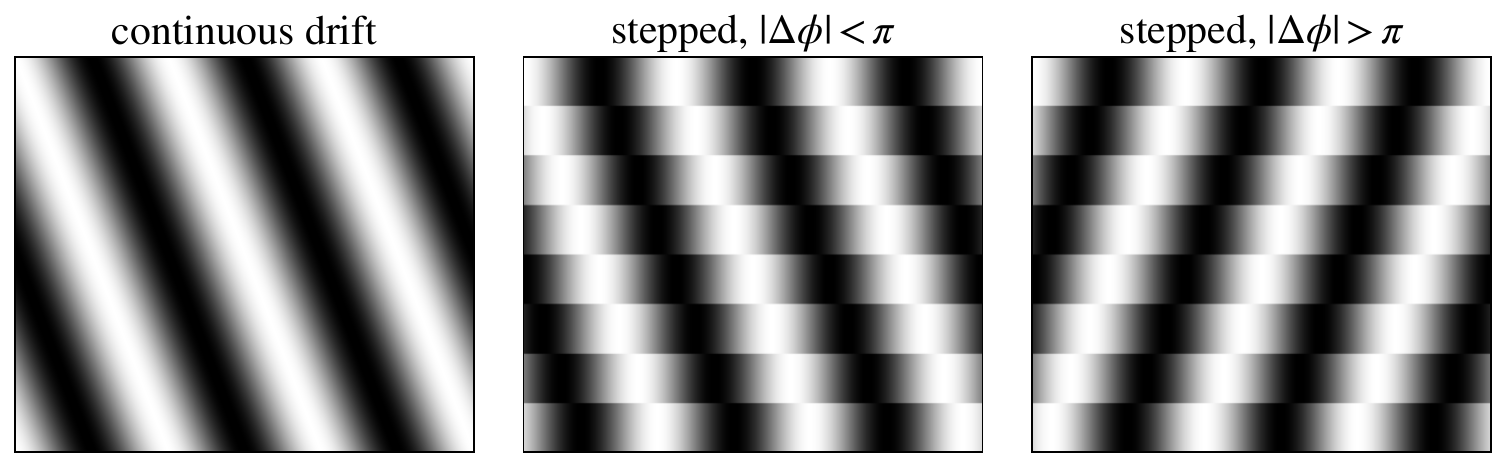}
        \caption{}
        \label{fig:phase-c}
    \end{subfigure}
    \caption{\textbf{(a)} Advancing the phase of a sinusoid by $|\Delta\phi| < \pi$ advances the wave forward on the x axis in 1D, giving forward motion.
    \textbf{(b)} Continuous drift discretized with stepped phase with $|\Delta\phi| \leq \pi$ gives a forward appearance of motion, while stepped phase with $|\Delta\phi| > \pi$ reverses the direction of apparent motion.
    }
    \label{fig:phase}
\end{figure}

\begin{figure}[htbp]
    \centering
    \includegraphics[width=\columnwidth]{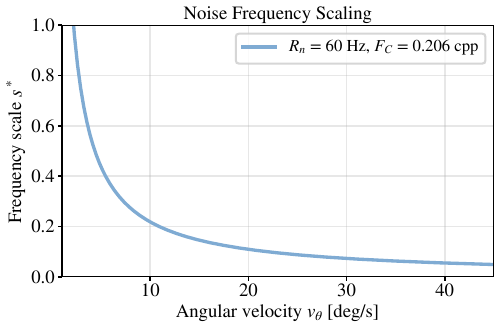}
    \caption{Noise frequency scaling predicted by Eq.~\ref{eq:sstar}, based on the center of the frequency synthesis band $[F_{L}, F_{H}]$.}
    \label{fig:sstar_ceiling_band}
\end{figure}

At higher velocities, frequencies of the synthesized Gabor noise have to be reduced to allow for the Gabor to appear as if it is translating via phase shifts.
In this section, we reveal how to render noise at such a desired frequency, by deriving a scale, $s^{*}$, that depends on the underlying frequency content, the render rate of the noise, and the angular velocity of on-screen content.

We first consider how the noise can be modulated between rendered frames.
The base Gabor noise of Section~\ref{sec:noise-synthesis} is synthesized from the rendered content.
When content is rendered at a reduced temporal sampling rate, there is no new content from which to synthesize new noise between frames.
Impulse locations could be randomly changed, but this would render new noise in a random way rather than carrying any specific motion.
The only free parameter in every \textit{displayed} frame is the phase, $\phi_{i}$, of each noise impulse.

To understand how motion is introduced into Gabor noise, it is useful to first consider the temporal behaviour of a single sinusoidal component. Let $h(x) = \cos(2 \, \pi \, f_{cpd} \, x)$ denote a sinusoidal function of angular position $x$, with spatial frequency
$f_{cpd}$ expressed in cycles per degree. If the sinusoidal pattern translates at a constant angular velocity $v_\theta$, its position becomes $x + v_{\theta} \, t$, giving
\begin{equation}
    h(x, t)
    = \cos\left[2\pi \, f_{cpd} \, ( x + v_{\theta} \, t)\right]
    = \cos\left(2\pi \, f_{cpd} \, x
    + 2\pi \, f_{cpd} \, v_{\theta} \, t\right)
\end{equation}
The second term represents a temporal phase progression with angular frequency
\begin{equation}
    \omega_{t} = 2\pi \, f_{cpd} \, v_{\theta}
\end{equation}
Using $\omega_{t} = 2\pi \, f_{t}$, the corresponding temporal frequency is therefore
\begin{equation}
    f_{t} = f_{cpd} \, v_\theta.
\label{eq:sinusoid-ft}
\end{equation}

This relationship applies directly to the individual sinusoidal components that constitute the synthesized Gabor noise. In our formulation, each impulse is associated with a sinusoidal component having spatial frequency $f_i$ (in cycles per pixel) and orientation $\theta_i$. The corresponding spatial frequency in angular units is $f_{i, cpd} = f_{i}  \cdot \eta$,
where $\eta$ denotes the display resolution in pixels per degree.
Consequently, translating a sinusoidal component by a displacement vector $\Delta x$ advances its phase by
\begin{equation}
    \Delta\phi_i
    = 2\pi \, f_i \langle \Delta x,\theta_{i} \rangle
\label{eq:phase-advance}
\end{equation}
where $\theta_i$ is the unit vector specifying the orientation of the sinusoidal component. Thus, the phase shift is simply the discrete counterpart of the continuous phase progression described above.

For continuous translation at angular velocity $v_\theta$, Eq.~\ref{eq:sinusoid-ft} gives the temporal frequency associated with each Gabor component as $f_{t} = f_{i,cpd}v_\theta$.
This temporal frequency determines how rapidly the phase of each sinusoidal component should shift over time.
Because the impulse locations are seeded deterministically (Section~\ref{sec:noise-synthesis}), the Gaussian envelopes remain fixed while the sinusoidal components drift beneath them.
This changes the apparent position of the spatial high frequency detail without translating the noise impulses; in peripheral vision, such coherent local drift can provide a perception of motion \cite{tariq2024towards}.

For each frequency band, we assign the sinusoidal components a frequency at the logarithmic midpoint of the band, $F_{C} = \sqrt{F_{L} F_{H}}$ (Section~\ref{sec:noise-synthesis}), and convert to cycles per degree (cpd).
\begin{equation}
    f_i = s F_{C},
    \qquad
    f_{i, cpd} = s \cdot \, F_{C} \cdot \eta
\end{equation}
where $s$ denotes the frequency scaling factor.

The noise is rendered at a discrete temporal rate $R_n$, which we refer to as the noise rate, so over one update interval $\Delta t = 1/R_n$ the phase advances by
\begin{equation}
    \Delta\phi_i
    = 2\pi \, f_{t} \, \Delta t
    = \frac{2\pi \, f_{t}}{R_{n}}
\label{eq:phase-step}
\end{equation}
The phase increment must be sufficiently small to avoid temporal aliasing.
If the phase advances by more than $\pi$ radians between successive noise updates, the sampled sinusoid violates the Nyquist criterion and may appear to reverse its direction of motion.
Such motion aliasing would be counterproductive to the goal of restoring motion perception and reducing the visibility of judder.
We therefore impose the constraint:
\begin{equation}
    |\Delta\phi_i| \leq \pi.
\label{eq:nyquist}
\end{equation}
In Figure~\ref{fig:phase-c}, we visualize how motion aliasing manifests from violating this constraint.
Importantly, this constraint is determined by the temporal sampling rate of the synthesized noise, $R_n$, rather than by the temporal rate of the underlying content that determines the judder alias (Eq.~\ref{eq:rho-judder}).
We therefore render the noise at the noise rate $R_n$, such that the constraint in Eq.~\ref{eq:nyquist} applies to each displayed phase step.
Expanding the constraint with Eq.~\ref{eq:phase-step},  $\frac{2\pi \, f_{t}}{R_n} \le \pi$, we can solve for the temporal frequency.
\begin{equation}
    f_{t} \le \frac{R_{n}}{2}
\label{eq:nyquist-ft}
\end{equation}
Given Eq.~\ref{eq:nyquist-ft}, the first alias copy of the phase stepped noise lies at a temporal frequency of at least $\frac{R_n}{2}$; we show below that this copy remains outside the boundary of visibility.
Substituting Eq.~\ref{eq:sinusoid-ft} expresses the constraint in terms of the spatial frequency and angular velocity:
\begin{equation}
     f_{i, cpd} \, v_{\theta}  \le \frac{R_{n}}{2}
\end{equation}
Since faster motion and higher spatial frequencies can both raise $f_{t}$, at a fixed angular velocity, the constraint becomes a ceiling on the spatial frequencies whose motion can be carried solely by the phase shift.
\begin{equation}
    f_{i, cpd} \le \frac{R_{n}}{2 \, v_{\theta}}
\end{equation}
The spatial frequency of the synthesized components is set by the frequency scale $s$ over the band (Section~\ref{sec:noise-synthesis}).
We seek a ceiling on the frequency scale.
Solving Eq.~\ref{eq:nyquist-ft} for $s$ with $f_{i, cpd} = s \cdot F_{C} \cdot \eta$, we can find the range of scale factors which we can use to scale the frequencies.
We select the maximum $s^{*}$ tenable, to minimize the visibility of generated noise.
\begin{equation}
s^{*} \le \frac{R_n}{2 \, v_{\theta} \, F_{C} \cdot \eta}
\label{eq:sstar}
\end{equation}
This final scale $s^{*}$ then multiplies by the synthesis band to synthesize the highest frequency noise at which phase shifts alone can still reasonably convey the motion of the synthesized noise at a given $v_{\theta}$.
At low velocities, the synthesis band itself can be phase shifted ($s^{*} \geq 1$), while faster motion requires scaling the band down so that each displayed phase step remains within Eq.~\ref{eq:nyquist}.
We clamp noise scales at $s^{*} = 1$ to avoid rendering undetectable noise.

We plot $s^{*}$ when using the center of the frequency band, $F_{C}$ (Figure~\ref{fig:sstar_ceiling_band}).
The scale constrains the frequency of the noise, but does not remove the content's aliasing copies; noise is composited on top of the rendered content, so aliasing copies from the rendering remain in place while the noise contributes its own signal.
As the noise is sampled at a higher rate, its own aliasing copy lies at a higher temporal frequency, outside of the boundary of visibility.
Therefore, the noise does not introduce its own judder; however,
as the frequency scales downwards, it may become resolvable and distracting in peripheral vision, introducing a potential opposition between image quality and reduction in judder visibility.
In the validation experiment (Section~\ref{sec:validation-experiment}), we measure how much the noise reduces perceived judder and explore this trade-off.

\begin{figure*}[tbp]
    \centering
    \includegraphics[width=\textwidth]{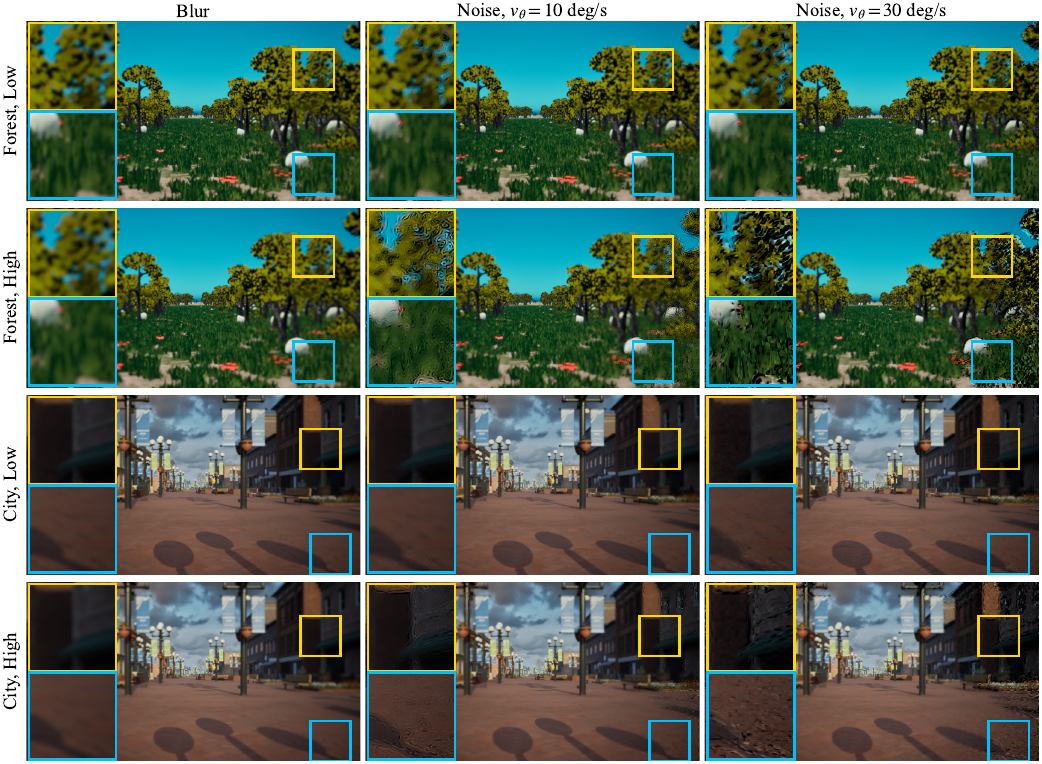}
    \caption{
    Example stimuli from the validation experiment, shown under all blur with low and high foveation levels, and noise over the low and high foveation levels at varying angular velocities.
    }
    \label{fig:validation-stimuli}
\end{figure*}

\subsection{Implementation \& Performance}
\label{subsec:implementation}

\begin{figure*}[tbp]
    \centering
    \includegraphics[width=\textwidth]{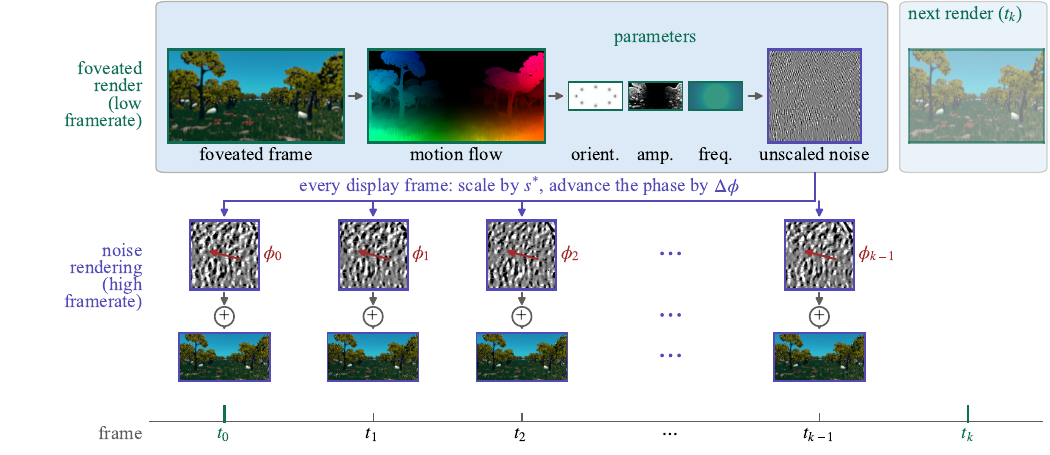}
    \caption{
    System overview.
    \textbf{Top:} at the reduced framerate, each foveated frame is rendered once.
    Kernel parameters (orientation, amplitude, and frequency band) and per-pixel motion flow are estimated from it to synthesize the unscaled noise.
    \textbf{Bottom:} at every display update, per-pixel adaptive scale $s^{*}$ (Eq.~\ref{eq:sstar}) is computed, which scales the synthesis band to prevent motion aliasing; the phase of the scaled noise advances along the local motion every displayed frame, and the result is composited onto the underlying frame.
    }
    \label{fig:system-diagram}
\end{figure*}

In our experimental setup, content is displayed at rate $R$, while the synthesized noise is updated at rate $R_{n}$, which we fix to $60$Hz in user testing.
The frequency scale is selected using Eq.~\ref{eq:sstar}.
Each displayed image combines the most recently rendered content at the low framerate with the updated noise, followed by the luminance correction described in Section~\ref{sec:noise-synthesis}.

Unreal Engine 5.6 with DirectX12 was used.
Foveated rendering was simulated by using a postprocess Gaussian blur in a compute shader.
The blurred output was read as an input texture to the noise synthesis shader to generate Gabor noise and phase shift the noise on a per-fragment basis.
Gaussian and Laplacian pyramids are efficiently constructed using GPU mipmaps of the swapchain image from Unreal's default rendering pipeline.
Unreal Engine generates motion vectors for camera-based motion as well as object based motion to support a variety of rendering effects.
The angular velocity of any particular point on screen is estimated from the motion vectors, which are used as inputs to our model to rescale ($s^{*}$) the synthesis band.
For optimization, amplitude, orientation, and frequency components are only computed when the underlying content is rendered.
The amplitude, orientation, and frequency components are stored between frames.
The noise synthesis and subsequent rescaling and phase shifting render extremely quickly, averaging $2.3$ms at 4K $(3840 \times 2160)$ spatial resolution, approximately $430$ frames per second on an NVIDIA RTX 5070.

While we simulate foveated rendering using Gaussian blurring, utilizing this in foveated renderers that perform true shading rate reductions will further improve performance.
Our full render pipeline is shown in Figure~\ref{fig:system-diagram}.
We used 32 impulses per cell for all user testing, as this produced the best balance for rendering budget and noise synthesis quality.

\section{Validation Experiment}
\label{sec:validation-experiment}

\begin{figure*}[tbp]
    \centering
    \includegraphics[width=\textwidth]{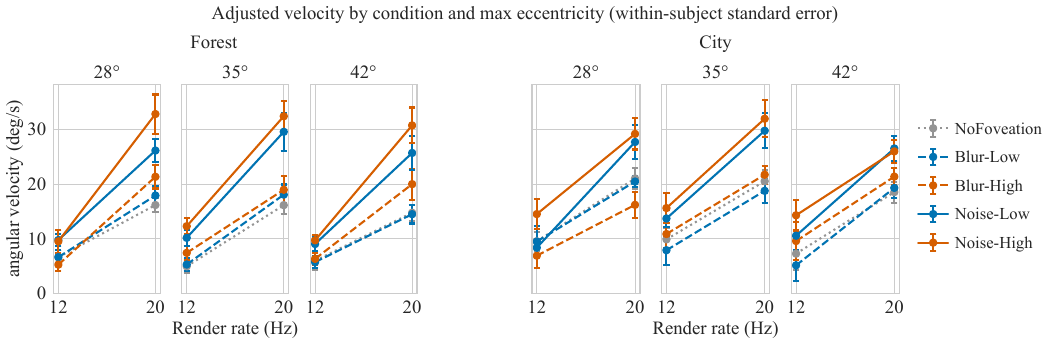}
    \caption{
    Adjusted angular velocities for the Forest (left) and City (right) scenes, per rendering condition, render rate, and maximum eccentricity.
    Higher velocities indicate stronger masking of judder.
    Error bars denote within-subject standard errors.
    }
    \label{fig:angular-velocity-results}
\end{figure*}

\begin{figure}[tbp]
    \centering
    \includegraphics[width=\linewidth]{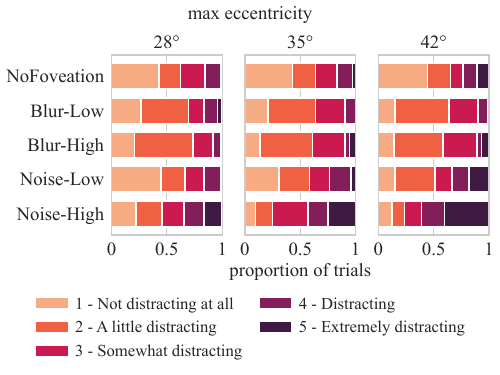}
    \caption{
    Distributions of distraction ratings per rendering condition and maximum eccentricity, pooled over scenes and render rates.
    Noise at the low foveation level is rated comparably to blur, while Noise-High incurs visibly more distraction, growing with eccentricity.
    }
    \label{fig:rating-distributions}
\end{figure}

Based on the promising exploratory results, we designed an IRB approved perceptual study (n=12) to validate to what extent our frequency scaling and phase shifts can mask motion judder.
At high angular velocities, our adaptive scaling (Eq.~\ref{eq:sstar}) lowers the noise frequency so that phase shifts can convey its motion within Eq.~\ref{eq:nyquist}, but the synthesized structure may become resolvable in peripheral vision.
Stronger foveation can further exacerbate this, as the rendered noise will already be constructed at a lower frequency band.
This introduces the trade-off that the study explores: \textit{even if motion judder is masked, what impact does the lower frequency noise that is being phase shifted have on subjective image quality?}

\textit{Setup.}
The same hardware as in the exploratory study was used.
To subtend a wider field of view, participants were seated 60cm from the screen and 6cm from its left edge, placing the right edge of the screen at an eccentricity of $42$ degrees.
Nasal eccentricity spans up to around $50$ degrees \cite{clark1990clinical}.
While the exploratory study moved stimuli at fixed eccentricities, here the whole visual field is in motion, so we assign eccentricity conditions based on the maximum visual angle subtended at the right edge of the screen.
Participants fixated a cross placed at different screen locations to target maximum right-edge eccentricities of $28, 35$, or $42$ degrees.
If the gaze fixation deviated from the cross, the screen would turn black until the gaze returned to the cross.
The central 10 degree radius of vision and all content to the left of the gaze point were grayed out to focus on peripheral motion judder.
As in the exploratory study, foveation was simulated with a Gaussian blur that increases with eccentricity.
The low foveation level was tuned per map to sit just below the visibility threshold, calibrated by three pilot participants naive to the purpose of the experiment, with the high foveation level set $50\%$ higher.
Noise was rendered at $60$Hz; while the monitor was capable of $120$Hz, we targeted $60$Hz to validate our method on more commonly used display types.
StelaCSF was used to estimate noise visibility at varying refresh rates (Figure~\ref{fig:gabor-sensitivity}), and the frequency scale was clamped to $s \geq 0.08$, the lowest estimated scale at $42$ degrees where the noise crosses the boundary of visibility.
Twelve students (Age 20-30, 9 male, 3 female) participated in the experiment.

\textit{Task.}
We tested five conditions: rendering without foveation, foveated rendering at two foveation levels (low and high), and our adaptive noise rendering at the same two levels.
Participants viewed two maps: Forest, a stylized path with grass and trees (low $\sigma = 0.12$, high $\sigma = 0.18$), and City (low $\sigma = 0.14$, high $\sigma = 0.21$), a downtown-style city block pathway.
To limit fatigue and eye strain reported during pilot testing, we test rendering the underlying scene content at two frame rates: ($20$Hz and $12$Hz).
We always render the noise over this at $60$Hz.
The noise was rendered with parameters that most closely match our blur values, linearly interpolated from \cite{tariq2022noise} Table 2.
Before the experiment, examples of smooth motion and juddery motion were shown.
In each trial, participants viewed a moving camera path under the tested rendering condition and adjusted the camera velocity up or down with the mouse scroll wheel until they perceived the motion as smooth.
They then rated the quality of the peripheral rendering based on how distracting the image is due to the rendering method on a scale of "$1$: Not distracting at all" to "$5$: Extremely distracting".
Example stimuli are shown in Figure~\ref{fig:validation-stimuli}.

\begin{table}[tbp]
    \centering
    \begin{tabular}{lccc}
    \toprule
    Comparison & Ratio & $p$ & $d_z$ \\
    \midrule
    Noise-High vs. Blur-High    & $1.62\times$ & \textbf{0.003} & $1.41$ \\
    Noise-High vs. NoFoveation  & $1.96\times$ & \textbf{0.003} & $1.39$ \\
    Noise-Low vs. Blur-Low      & $1.56\times$ & \textbf{0.004} & $1.28$ \\
    Noise-Low vs. NoFoveation   & $1.66\times$ & \textbf{0.007} & $1.14$ \\
    Blur-High vs. NoFoveation   & $1.21\times$ & $0.108$ & $0.62$ \\
    Blur-Low vs. NoFoveation    & $1.06\times$ & $0.302$ & $0.31$ \\
    \bottomrule
    \end{tabular}
    \caption{
    Comparisons of adjusted velocity rates.
    Paired t-tests on participant velocities, pooled over scenes and eccentricities, and the Holm-corrected $p$ values are shown.
    Effect sizes ($d_{z}$) are shown in the last column.
    We note large increases in comfortable perceived velocities using our adaptive noise scaling as compared to foveation or no foveation.}
    \label{tab:velocity-comparisons}

\end{table}

\textit{Results.}
The adjusted angular velocities are shown in Figure~\ref{fig:angular-velocity-results}.
The plots show the velocity participants adjusted to before perceiving the motion as juddery.
Error bars denote standard errors.
For significance reporting, paired t-tests on within-subject velocities, pooled over scenes and eccentricities with Holm corrections were performed.
Full results are shown in Table~\ref{tab:velocity-comparisons}.
Consistently, the results indicate that our phase shifts allow viewers to sustain markedly higher angular velocities of camera motion at low framerates than foveation and the no-foveation conditions.
Pooled over scenes and eccentricities, our adaptive noise raised the tolerated velocity $1.96\times$ over full resolution rendering (no foveation) at the high foveation level and $1.66\times$ at the low level.
Noise methods also raised angular velocities before motion was perceived as juddery $1.62\times$ and $1.56\times$ over blur at the same foveation levels.
This is consistent with the frequency-domain view of Section~\ref{sec:method}; at the velocities where judder becomes objectionable, the first aliasing copy lies at spatial frequencies well below the foveation cut-off, so removing high frequency content neither conceals the alias nor removes the low frequency content energy from which it arises.
Masking requires occupying the alias band with new, smoothly moving energy, which we accomplish via phase shifts of the noise.

Figure~\ref{fig:rating-distributions} shows the distributions of distraction ratings.
Ratings were taken after the motion was judged to be smooth, and participants were instructed to judge only the image quality, separately from the smoothness of motion.
At the low foveation level, ratings for the noise did not differ from matched blur (Wilcoxon signed-rank, $p = 0.83$), despite the $1.56\times$ velocity benefit.
At the high level of foveation with noise, ratings increased $+0.67$ over matched blur ($p = 0.004$, Holm-corrected), indicating the noise became resolvable and distracting to users.
Within the noise conditions, ratings rose with the adjusted velocity (median within-participant Spearman $\rho = 0.15$, $p = 0.007$), matching the behaviour our model predicts: as velocity grows, Eq.~\ref{eq:sstar} lowers the synthesized band toward resolvable frequencies.

The distinct gap in preference between the two noise conditions does not stem from the noise synthesis itself, since the synthesis between them is identical.
The difference arises due to the amount of removed spatial spectrum they must recover, caused by the difference in foveation levels.
At low foveation levels that sit right below detectability, there is little missing of the spectrum; the synthesized noise band lies well above resolvable frequencies, allowing more room to scale the noise to lower frequencies to prevent motion aliasing before the noise becomes spatially resolvable.
However, at high levels of foveation, where more spatial content has been removed, the synthesized noise has to replace a larger loss in spectrum, and renders closer to the resolvable spectrum.
This, combined with the motion aliasing constraint, results in noise that, at high angular velocities, becomes more resolvable and objectionable to view, despite masking judder well.
This points to an interesting conclusion about the trade-off with noise synthesis and judder, as masking judder by using noise to restore spatiotemporal energy  does not necessarily have to cost image quality.
As long as the synthesized band remains above resolvable frequencies, the velocity benefit can come without a negative perceptual impact.

\section{Limitations \& Future Work}
While we have found promising results on mitigating motion judder, our approach is not without limitations.
Firstly, we specify only how to synthesize and advance the noise via phase shifts along the direction of motion for a chosen level of foveation, velocity, and rendering rate.
Although we note a trade-off in judder mitigation and resulting image quality trade-offs, our experiments characterize the trade-off between judder reduction and the resulting image quality from lowering the noise frequency, but we do not develop a general model to automatically tune the configuration to be optimal.
Future work could develop a model that jointly selects the content rate, foveation level, and noise parameters according to parameters such as viewing conditions, scene content, and the available rendering budget.
Such a model would complement the frequency constraint derived here (Section~\ref{sec:method}) by predicting the perceptual consequences of the selected configuration, and providing a rendering setup that minimizes judder while maximizing image quality.
Recent works have begun to explore how generative models can simulate human peripheral vision, or generate content designed for it \cite{kerkouri2026matched, raina2026generating, chao2026foveated}.
There is also some potential to explore adjusting resolvable noise perturbations by directly modifying the input via a learned model \cite{biparva2026tutor}.
Although we implement some caching to optimize the phase shifts of Gabor noise frame-to-frame, more optimizations could be done to improve the noise generation efficiency even further.

Low frame rates, and the judder they introduce, can aggravate cybersickness, while adding high frequency noise that provides countervection cues has been shown to reduce it \cite{groth2026gaze}, although only at the full temporal resolution of the display.
As our noise supplies such high frequency content and restores the perception of smooth motion at reduced rendering rates, it may also help mitigate cybersickness in VR, which future work could evaluate.

Lastly, our user study only measures the trade-off between judder masking and image quality at the highest angular velocity at which motion appears smooth; rather, another study could stop the adjustment method when image quality begins to degrade, and measure the velocity, to fine-tune specifically how much judder mitigation our method can get while maintaining the same image quality, irrespective of foveation levels or frequency bands.

\section{Conclusion}
Foveated rendering saves computation by reducing spatial resolution in the periphery, and prior work has shown that inexpensive procedural noise can restore the lost detail \cite{tariq2022noise} and the motion energy it carried \cite{tariq2024towards}.
Unlike these prior works that focus on foveated rendering at the full framerate, our method reduces it further to limit the computational cost.
While limiting framerate introduces visible judder, we show that synthesizing \textit{only} noise at higher frame rates can vastly limit the visibility of judder, and advance the limits of foveation further into the temporal domain.
Because the noise parameters are estimated only when the content is rendered, each intermediate frame costs only the phase shift and rescaling of the noise, enabling much faster performance than rendering the full frame.
In our validation experiment, phase-shifted noise allowed viewers to tolerate substantially higher angular velocities before perceiving judder than both foveated and full-resolution rendering at the same reduced frame rates, and at a foveation level just below visibility, it did so without increasing distraction relative to blur alone.

\bibliographystyle{eg-alpha-doi}
\bibliography{ville-bib}

\end{document}